\documentstyle[12pt,a4]{article}

\newcommand{\news}{\setcounter{equation}{0}}

\def\eqn{\begin{equation}}
\def\eeqn{\end{equation}}
\def\arr{\begin{array}}
\def\earr{\end{array}}
\def\a{\alpha}
\def\e{\epsilon}

\def\tr{f^{-1}dr^{2}+r^{2}d\Omega^{2}}

\font\mybb=msbm10 at 12pt
\def\bb#1{\hbox{\mybb#1}}
\def\bR {\bb{R}}

\begin{document}

\title{Black composite M-branes}

\author{Miguel S. Costa\thanks{email address:
M.S.Costa@damtp.cam.ac.uk}\\\\
D.A.M.T.P.\\
University of Cambridge\\
Silver Street\\
Cambridge CB3 9EW\\
England}

\date{October 1996}

\maketitle

\begin{abstract}
We generalise all the known supersymmetric composite M-branes to the
corresponding black configurations. Thermodynamical formulae is
written by using the simple rules to construct these black branes.
\end{abstract}

\newpage

\section{Introduction}
\news

In the past year string theory has addressed a longstanding problem in
black hole physics, the statistical origin of black hole
entropy. In particular, by interpreting certain BPS saturated
black holes carrying R-R charges as Dirichlet branes \cite{pol}, it
has been possible to provide a microscopic explanation of the
Bekenstein-Hawking entropy $[2-6]$. The existence of supersymmetric
black holes with finite horizon area in four \cite{cy,ct1,ct2} and
five \cite{ts2,sv} dimensions has been crucial in these
calculations. The entropy and Hawking temperature of near extremal
black holes have also been successfully calculated
\cite{cm,hs,blpsv}.

During the same period M-theory has been a subject of intensive
research. The effective field theory limit of this conjectured theory
is $11D$ supergravity. Remarkably, it seems that the different
superstring theories and corresponding duality symmetries have an
eleven-dimensional origin $[13-16]$. A related issue is the
eleven-dimensional origin of the supersymmetric brane solutions of
type IIA 
superstring theory \cite{t}. Therefore it is important to study the
supersymmetric brane solutions of $11D$ supergravity, not only to try
to understand what M-theory may or may not be, but also to find new
supersymmetric brane solutions of superstring theories. The first step
in this direction was given by Townsend and Papadopoulos \cite{pt} who
reinterpreted some of the brane solutions found by G\"{u}ven
\cite{g} as orthogonal intersections of basic membranes
\cite{ds}. They also found the corresponding magnetic duals
interpreted as orthogonal intersections of the basic 5-branes
\cite{g}. Tseytlin \cite{ts} (see also \cite{kt,gkt}) has extended
this work and formulated a general rule, the harmonic function rule,
to construct orthogonal intersecting branes with the basic 2 and 5
branes as its constituents. Some configurations can be boosted along a
common string to all branes \cite{ts} and/or superposed with a Ka\l
u\.{z}a-Klein (KK) monopole \cite{c} yielding, upon dimensional
reduction to $D=10$, brane solutions with charges arising from the R-R
2-form field strength of type IIA superstring theory. Intersecting
configurations with the $\frac{1}{2}$ supersymmetric $(2\subset
5)$-brane \cite{ilpt} among its constituents were found in \cite{c}.

The black hole solutions of lower
dimensional theories may be interpreted as reductions of extreme and
non-extreme 
composite M-branes. One of the challenges of M-theory is to provide
the statistical counting of the 
Bekenstein-Hawking entropy of these black holes from an eleven
dimensional perspective
$[22,26-28]$. It is therefore important to have 
a complete classification of the black composite M-branes and
corresponding black hole solutions of lower dimensional theories
\cite{dlp,ct}. The aim of this paper is to carry on such a program by
considering the non-extreme configurations associated with the new extreme
solutions found in \cite{c}.

This paper is organised as follows. In section two we start by
presenting the black $(2\subset 5)$-brane solution and by reviewing
the composite M-brane rules. We then state the rules to construct
the black composite M-branes. In
section three we will give 
the eleven-dimensional interpretation of black hole solutions in
$D$-dimensional spacetime for $4\le D\le 10$. Finally, in section four we
discuss  the thermodynamics of these
black hole solutions.

\section{Black M-branes from extreme M-branes}
\news

Through this paper we will be considering bosonic solutions of $11D$
supergravity whose action is
\eqn
I_{11}=\frac{1}{2}\int
d^{11}x\sqrt{-g}\left[R-\frac{1}{2.4!}{{\cal F}}^{2}\right]+
\frac{1}{12}\int {\cal F}\wedge {\cal F}\wedge {\cal A},
\eeqn
where ${\cal F}=d{\cal A}$ and ${\cal A}$ is a 3-form
field. In this section we will describe the
method to construct the most general black composite M-branes from the
corresponding extreme configurations. We will start by presenting the
non-extreme generalisation of the $(2\subset 5)$-brane
solution. The composite M-brane rules needed for our purposes will
then be reviewed. We will follow the
notation of \cite{c}.

\subsection{Black $(2\subset 5)$-brane}

Nearly all of the new supersymmetric configurations found in \cite{c}
had $(2\subset 5)$-branes among its constituents. Even thought the
corresponding Chern-Simons term in the action (2.1) is non-vanishing
those configurations solve the equations of motion that follow from
this action. It turns out
that there is also a generalisation of the $\frac{1}{2}$
supersymmetric $(2\subset 5)$-brane to the corresponding non-extreme
configuration. This anisotropic black $p$-brane is described by
($5\le p\le 7$)
\eqn
\arr{l}
\arr{ll}
ds^{2}= & \left(H\tilde{H}\right)^{\frac{1}{3}}\left[ H^{-1}\left(
-fdt^{2}+dy_{1}^{2}+dy_{2}^{2}\right)
\phantom{\tilde{H}^{-1}}\right.
\\\\ & \left. + \tilde{H}^{-1}\left( dy_{3}^{2}+dy_{4}^{2}+
dy_{5}^{2}\right)+dy_{6}^{2}+...+dy_{p}^{2}+\tr_{9-p}\right],
\earr
\\\\ {\cal F}_{(2\subset 5)}={\cal F}_{(2)}+{\cal F}_{(5)}
-\tan{\zeta}\ d\left(\tilde{H}^{-1}-1\right)\wedge \xi ,
\earr
\eeqn
where $*{\cal F}_{(2)}=Q(\e_{9-p}\wedge\eta)$, ${\cal
  F}_{(5)}=P(\mu\wedge\e_{9-p})$, $f=1-\frac{\mu}{r^{8-p}}$,
$H=1+\frac{\a}{r^{8-p}}$ and 
$\tilde{H}=\sin^2{\zeta}+\cos^2{\zeta}\ H$.  The dual operation in the
expression for ${\cal F}_{(2)}$
is defined for the metric with $\cos{\zeta}=0$.
 $\e_{9-p}$ is the unit $(9-p)$-sphere volume form. $\mu
=dy_{6}\wedge ...\wedge dy_{p}$ and $\eta
=dy_{3}\wedge dy_{4} \wedge dy_{5}\wedge\mu$ are the volume forms on
the relative transverse spaces of the 5-brane $\left({\cal
    M}_{(5)}^{p-5}\right)$
and 2-brane $\left({\cal M}_{(2)}^{p-2}\right)$, respectively. $\xi
= dy_{3}\wedge dy_{4}\wedge dy_{5}$ is the volume form on the space
${\cal M}_{(5/2)}^{3}$ spanned by the vectors that are tangent to the
5-brane's worldvolume but are orthogonal to the 2-brane's
worldvolume. If $\cos{\zeta}=0$ we obtain the
black 2-brane solution (with $p-2$ extra flat directions) and if
$\sin{\zeta}=0$ the black 5-brane solution (with $p-5$ extra flat
directions).
If $\mu=0$ we obtain the extreme
supersymmetric configuration.\footnote{We use the same Greek letter
  to denote the volume form on the space ${\cal M}_{(5)}^{p-5}$ and
  the non-extremality parameter. It should be obvious from
  the context which one we mean.}
The electric charge is
defined by $Q=\frac{1}{V_{(2)}^{p-2}}\int _{\Sigma}*{\cal F}$,
where $V_{(2)}^{p-2}$ is the volume of ${\cal  M}_{(2)}^{p-2}$ and 
$\Sigma =S^{9-p}\times {\cal M}_{(2)}^{p-2}$ is an asymptotic
spacelike hypersurface.\footnote{We remark
that from the 4-form field equation, the conserved electric charge is
in fact given by $Q=\frac{1}{V_{(2)}^{p-2}}\int_{\Sigma}\left(\star{\cal
    F}+\frac{1}{2}{\cal A}\wedge{\cal F} \right )$. In the cases where
the Chern-Simons term in this integral is non-vanishing as it is the case
for this solution, we will always
choose a gauge such that it vanishes at spatial infinity and therefore
the charge is still given by the expression in the text.} The magnetic
charge is 
defined by $P=\frac{1}{V_{(5)}^{p-5}}\int _{\Sigma}{\cal F}$
with $\Sigma ={\cal M}_{(5)}^{p-5}\times S^{9-p}$ and $V_{(5)}^{p-5}$
the volume of ${\cal  M}_{(5)}^{p-5}$. These charges are given by
\eqn
\arr{l}
\left(\frac{Q}{(8-p)A_{9-p}}\right)^{2}=\a\left(
  \mu+\a\right)\sin^{2}{\zeta}\ ,
\\\\ 
\left(\frac{P}{(8-p)A_{9-p}}\right)^{2}=\a\left(
  \mu+\a\right)\cos^{2}{\zeta}.
\earr
\eeqn
where $A_{9-p}$ is the volume of the unit $(9-p)$-sphere.

Two remarks are in order. Firstly, this solution has a non-vanishing
Chern-Simons term. This fact slightly complicates the corresponding
field equations. It also shows that the black $p$-brane solutions found in
\cite{dlp} are not the most general solutions. In fact, in \cite{dlp}
(and in the related work \cite{lpx})
a $D$-dimensional Lagrangian derived by reduction of the $11D$
supergravity bosonic Lagrangian was used \cite{lp}. The
Chern-Simons term in the action (2.1), as well as the Chern-Simons
type modifications to the field strengths due to the dimensional
reduction process were then assumed to vanish. Secondly, the existence of
this solution may be argued as a consequence of the $SL(2,\bR)$
electromagnetic duality of $D=8$, $N=2$ supergravity. This can be done
by starting with the (electrically or magnetically charged) black
membrane solution of the latter theory and 
by following the same steps that led to the construction of the
extreme $(2\subset 5)$-brane solution \cite{ilpt}.

\subsection{Extreme M-brane rules}

As mentioned at the end of the previous subsection, black $p$-brane
solutions can be obtained by performing a consistent 
truncation of the $11D$ supergravity bosonic Lagrangian to
$D$-dimensions \cite{dlp}. Not only this approach as the disadvantage
of enforcing the constraint that the Chern-Simons terms have to
vanish, but also it is much more complicated due to the presence of a
much larger number of fields. The eleven-dimensional approach is
simpler, does not impose any constraint and provides an immediate
interpretation of the solutions in terms of compositions of M-branes. 
Because the non-extreme composite M-branes can be easily obtained from
the corresponding one-centered extreme solutions we will briefly
review the composite M-brane rules. 

The rules to construct configurations of $N$ $\frac{1}{2}$
supersymmetric M-branes are:

\begin{description}
\item[(i)]
To each basic $q_{i}$-brane we assign an harmonic function $H_{i}$ on the
overall transverse space. If the coordinate $y$ belongs to several
constituents $q_{i}$-branes ($q_{1},...,q_{n}$) then its contribution to
the metric written in the conformal frame where the overall transverse
space is 'free' is $H_{1}^{-1}...H_{n}^{-1}dy^{2}$
\cite{ts}.\footnote{There is an extra minus sign if $y$ is the time
  coordinate $t$.} The contribution to the conformal factor of the
i-th $q_{i}$-brane is $H_{i}^{\frac{q_{i}+1}{9}}$. The 4-form field
strength is given by
\eqn
{\cal F}=\sum _{i=1}^{N} {\cal F}_{(q,i)},
\eeqn
where
\eqn
*{\cal F}_{(2,i)}=Q_{i}(\e_{9-p}\wedge\eta_{i}),\ \ 
{\cal F}_{(5,i)}=P_{i}(\mu_{i}\wedge \e_{9-p}),
\eeqn
whether the i-th brane is a 2 or 5-brane, respectively. In the former
case, $\eta_{i}$ is the volume form on the i-th 2-brane relative
transverse space ${\cal M}_{(2,i)}^{p-2}$ and the electric charge is defined
by $Q_{i}=\frac{1}{V_{(2,i)}^{p-2}}\int _{\Sigma}*{\cal F}_{(2,i)}$ with
$\Sigma =S^{9-p}\times {\cal M}_{(2,i)}^{p-2}$ and $V_{(2,i)}^{p-2}$
the volume of ${\cal  M}_{(2,i)}^{p-2}$. In the latter case, $\mu_{i}$
is the volume form on ${\cal M}_{(5,i)}^{p-5}$ and the magnetic charge is
defined by $P_{i}=\frac{1}{V_{(5,i)}^{p-5}}\int _{\Sigma}{\cal F}_{(5,i)}$
with $\Sigma ={\cal M}_{(5,i)}^{p-5}\times S^{9-p}$ and $V_{(5,i)}^{p-5}$
the volume of ${\cal  M}_{(5,i)}^{p-5}$. Both the electric and
magnetic charges are given by
\eqn
\frac{Q_{i}}{(8-p)A_{9-p}},\ \frac{P_{i}}{(8-p)A_{9-p}}=\pm\a_{i}.
\eeqn
\

\item[(ii)]
$q$-branes of the same type can intersect orthogonally over
($q-2$)-branes \cite{pt}. A 2-brane can intersect orthogonally a
5-brane over a string \cite{st,t2}.
\

\item[(iii)]
If there are ($2\subset 5$)-branes among the $N$ intersecting
constituents then in the cases where these branes reduce to the basic 2
or 5-branes the resulting configuration should be compactible with the
previous rules (i) and (ii) \cite{c}. The 4-form field strength is
given by
\eqn
{\cal F}=\sum _{i=1}^{N} {\cal F}_{(q,i)},
\eeqn
where ${\cal F}_{(q,i)}$ is given either by one of the expressions in
(2.5) or by\footnote{The sign of the third term in (2.8) depends on
  how we distribute the constituent M-branes. We will always make the
  minus sign choices.}
\eqn
{\cal F}_{(2\subset 5,i)}={\cal F}_{(2,i)}+{\cal F}_{(5,i)}-
\tan{\zeta_i}\ d\left(\tilde{H_i}^{-1}-1\right)\wedge \xi_i .
\eeqn
The dual operations are defined for $\cos{\zeta_i}=0$ and $\xi_{i}$
is the volume form on the space ${\cal M}_{(5/2,i)}^{3}$. The
electromagnetic charges are defined as before. They are given by
\eqn
\frac{Q_{i}}{(8-p)A_{9-p}}=\a_{i}\sin{\zeta_{i}}\ ,\ \ 
\frac{P_{i}}{(8-p)A_{9-p}}=\a_{i}\cos{\zeta_{i}}\ ,
\eeqn
where $\cos{\zeta_{i}}=0$ or $\sin{\zeta_{i}}=0$ if the i-th brane is
a basic 2 or 5-brane, respectively.

\end{description}

Some configurations can be boosted along a common string to all branes
\cite{ts} and/or superposed with a KK monopole \cite{c}. For black
configurations we will consider a variation of this fourth rule. The
last rule is concerned with the amount of preserved supersymmetry and
it will not be used here.

\subsection{Black composite M-brane rules}

One-center supersymmetric compositions of M-branes can be used to
construct the corresponding non-extreme configurations. These
configurations may be seen as anisotropic black $p$-branes and can be
built according to the following rules:

\begin{description}
\item[(i)]
Perform the following replacements in the metric \cite{dlp,ct}
\eqn
\arr{l}
dt^{2}\rightarrow fdt^{2}
\\\\
dx^{i}dx_{i}\rightarrow \tr_{9-p},
\earr
\eeqn
where $i=1,...,10-p$ and $f=1-\frac{\mu}{r^{8-p}}$ with
$r^{2}=x^{i}x_{i}$.
\

\item[(ii)]
The 4-form field strength ansatz remains the same with the charge
redefinitions
\eqn
\arr{l}
\left(\frac{Q_{i}}{(8-p)A_{9-p}}\right)^{2}=\a_{i}^{2}\sin^{2}{\zeta_{i}}
\rightarrow \a_{i}\left(\mu+\a_{i}\right)\sin^{2}{\zeta_{i}}\ ,
\\\\ 
\left(\frac{P_{i}}{(8-p)A_{9-p}}\right)^{2}=\a_{i}^{2}\cos^{2}{\zeta_{i}}
\rightarrow \a_{i}\left(\mu+\a_{i}\right)\cos^{2}{\zeta_{i}},
\earr
\eeqn
where $\cos{\zeta_{i}}=0$ or $\sin{\zeta_{i}}=0$ when the
corresponding i-th brane is a 2 or 5-brane, respectively.
\end{description}

Notice that rule (ii) above has been given in \cite{dlp} for the cases
where $\cos{\zeta_i}=0$ or $\sin{\zeta_i}=0$. As explained before the
more general cases considered here allow configurations with
non-vanishing Chern-Simons terms.

In some configurations we can add KK charges yielding, upon
dimensional reduction to $D=10$, branes with charges arising from the
R-R 2-form field strength of type IIA superstring theory.
Correspondingly, the Ka\l u\.{z}a-Klein reduction of $11D$ supergravity
yields $N=2A$, $D=10$ supergravity. The reduction of the bosonic
fields is performed by writing
\eqn
\arr{l}
ds^{2}=g_{MN}dx^{M}dx^{N}=e^{-\frac{2}{3}\phi
}g_{mn}dx^{m}dx^{n}+e^{\frac{4}{3}\phi
}\left( dx^{10}-{\cal A}_{m}dx^{m}\right)^{2},\\\\
{\cal A}_{MNP}={\cal A}_{mnp},\ {\cal A}_{MN10}={\cal B}_{mn},\ \ with\
\ M,N,P=0,...,9.
\earr
\eeqn
Unless stated capital letters range from 0 to 10 and in the lower case
from 0 to 9. The rule to generate black branes with KK charges is:

\begin{description}
\item[(iii)]
If the corresponding extreme configuration has a common string to all
constituent M-branes, say along $y$, then a KK electric charge can be
added by applying the boost transformation \cite{ct}
\eqn
\arr{l}
t\rightarrow\left( 1+\frac{\a}{\mu}\right)^{\frac{1}{2}}t\mp 
\left(\frac{\a}{\mu}\right)^{\frac{1}{2}}y,
\\\\
y\rightarrow\mp \left(\frac{\a}{\mu}\right)^{\frac{1}{2}}t+
\left( 1+\frac{\a}{\mu}\right)^{\frac{1}{2}}y,
\earr
\eeqn
where the $\mp$ sign choice will correspond to positive or negative KK
charge, respectively. The $g_{tt}$ and $g_{yy}$ elements of the black $p$-brane
are then transformed to
\eqn
-fdt^{2}+dy\rightarrow -H^{-1}fdt^{2}+
H\left(dy\mp\frac{\sqrt{\a\left(\mu+\a\right)}}{r^{8-p}+\a}dt\right)^{2},
\eeqn
where $H=1+\frac{\a}{r^{8-p}}$.
Compactifying along the $y$ direction we have a black composite brane
with a 0-brane among its constituents.

If the overall transverse space has dimension bigger than
three, a magnetic monopole can be added by making all the harmonic
functions\footnote{We refer to $H_{i}$ as harmonic functions
  even thought they are no longer harmonic functions of the non-extreme
  metrics.} to depend only on three of this space coordinates and
performing the substitution (we start with an anisotropic black
$p$-brane)
\eqn
\arr{l}
\tr_{9-p}\rightarrow dy_{p+1}^{2}+...+dy_{6}^{2} 
\\\\ +H^{-1}\left( dy_{7}\pm \sqrt{\a\left(\mu+\a\right)}
  \cos{\theta}d\phi\right)^{2}+H\left(\tr_{2}\right),
\earr
\eeqn
where $H=1+\frac{\a}{r}$, $f$ is transformed to
$f=1-\frac{\mu}{r}$ and the $\pm$ sign choice will correspond to
positive or negative KK charge, respectively. Compactifying along
the $y_{7}$ direction we have a black composite brane with a 6-brane
among its constituents.

There are cases where we can add both electric and magnetic KK
charges and a further compactification is required.

\end{description}

We remark that this rule, for both electric and magnetic KK charges,
may be recasted  in the corresponding dimensionally reduced theories
in a form similar to the rule (ii) above (for $\cos{\zeta_i}=0$ or
$\sin{\zeta_i}=0$).

\section{Black holes in lower dimensions}
\news

\begin{table}
\hbox{\hskip1.2in
\begin{tabular}{c}
TABLE I
\\
\\
\begin{tabular}{|c|c|c|} \hline
        &       & Black composite M-brane \\ \hline \hline
$D=4$   & $n=4$ & $2\perp 2\perp 5\perp 5$ \\
        &       & $5\perp 5\perp 5 + boost$ \\ 
        &       & $2\perp 2\perp 2+KK\ monopole$ \\
        &       & $2\perp 5 + boost + KK \ monopole$ \\ \cline {2-3}
        & $n=3$ & $(2\subset 5)\perp (2\subset 5)\perp (2\subset 5)$\\
        &       & $(2\subset 5)\perp 5 + boost$ \\ 
        &       & $(2\subset 5)\perp 2 + KK \ monopole$ \\
        &       & $(2\subset 5)+boost + KK \ monopole$   \\ \hline 
$D=5$   & $n=3$ & $2\perp 2\perp 2$ \\
        &       & $2\perp 5 + boost$ \\ \cline {2-3}
        & $n=2$ & $(2\subset 5)\perp 2$ \\ \hline 
$D=6$   & $n=2$ & $(2\subset 5) + boost$ \\
        &       & $2\perp 2$ \\ \hline
$D=7$   & $n=2$ & $2\perp 2$ \\
        &       & $2 + boost$ \\ \hline
$D=8,9$ & $n=2$ & $2 + boost$ \\ \hline
$D=10$  & $n=1$ & $boost$ \\ \hline
\end{tabular}
\end{tabular}\hfill}
\end{table}

In table I we present the most general black M-brane configurations
that reduce to black holes in lower dimensions after dimensional
reduction. Consider first black holes in four dimensions with $n=4$.
The black $2\perp 2\perp 5\perp 5$ and $5\perp 
5\perp 5 + boost$ branes were presented in \cite{ct} and the other two
follow from the KK monopole rule stated in section 2.3. All these
solutions have been obtained from a consistent reduction to four
dimensions of the $11D$ theory \cite{dlp}. Here we present the corresponding
eleven dimensional interpretation. The most general $D=4$ configurations
with $n=3$ are also presented. In order to avoid an excessive lengthy
exposition and since it is straightforward to construct these
solutions by using the rules of the previous section we will just
present the black $(2\subset 5)\perp (2\subset 5)\perp (2\subset 5)$
  brane solution. It is described by 
\eqn
\arr{l}
\arr{ll}
ds^{2}= & \left({\displaystyle \prod_{i=1}^{3}}\left( H_{i}\tilde{H}_{i}
\right)^{\frac{1}{3}}\right) \left[ -\left( H_{1}H_{2}H_{3}\right)
^{-1}fdt^{2}+(H_{1}\tilde{H}_{3})^{-1}dy_{1}^{2}\right.
\\\\ & +(H_{1}\tilde{H}_{2})^{-1}dy_{2}^{2}+
(H_{2}\tilde{H}_{1})^{-1}dy_{3}^{2}+
(H_{2}\tilde{H}_{3})^{-1}dy_{4}^{2}+
(H_{3}\tilde{H}_{2})^{-1}dy_{5}^{2} 
\\\\ & \left. +(H_{3}\tilde{H}_{1})^{-1}dy_{6}^{2}+
(\tilde{H}_{1}\tilde{H}_{2}\tilde{H}_{3})^{-1}dy_{7}^{2}+\tr_{2}\right],
\earr
\\\\
{\cal F}={\displaystyle \sum_{i=1}^{3}}
\left( {\cal F}_{(2,i)}+{\cal F}_{(5,i)}
-\tan{\zeta_i}\ d\left(\tilde{H_i}^{-1}-1\right)\wedge \xi_i\right),
\earr
\eeqn
where $*{\cal F}_{(2,i)}=Q_{i}(\e_{2}\wedge\eta_{i})$ and
${\cal F}_{(5,i)}=P_{i}(\mu_{i}\wedge\e_{2})$.
The electric and magnetic charges are given by
\eqn
\arr{l}
\left(\frac{Q_{i}}{A_{2}}\right)^{2}=
\a_{i}\left(\mu+\a_{i}\right)\sin^{2}{\zeta_{i}},
\\\\ 
\left(\frac{P_{i}}{A_{2}}\right)^{2}=
\a_{i}\left(\mu+\a_{i}\right)\cos^{2}{\zeta_{i}},
\earr
\eeqn
with $i=1,2,3$.
Four-dimensional solutions with $n\le 2$ can be obtained by allowing some of
the charges of the corresponding configurations in table I to vanish. 

The eleven-dimensional interpretation of black hole solutions in
$D$ spacetime dimensions for
$5\le D\le 10$ is also described in table I. There are solutions that have
the same eleven-dimensional interpretation but reduce to black hole
solutions in different spacetime dimensions. The reason is that the
corresponding harmonic functions have a different dependence on the
overall transverse space. Solutions with $n=1$ can be obtained by
allowing all charges but one to vanish.

\section{Thermodynamics}
\news

Remarkably, toroidal compactification of all black $p$-branes of
M-theory along the branes spatial directions yields the following
$(11-p)$-dimensional Einstein's frame metric
\eqn
\arr{l}
ds^{2}=-\lambda^{3-D}(r)fdt^{2}+\lambda (r)\left[ \tr_{D-2}\right] ,
\\\\
\lambda (r)=
\left( {\displaystyle \prod_{i=1}^{n}}H_{i}(r)\right)^{\frac{1}{D-2}}.
\earr
\eeqn
Notice that for configurations involving the eleven-dimensional
$(2\subset 5)$-brane the
metric still reduces to this simple form. However, the dilaton field
as well as the field strengths will be much more complicated than in
the other cases.

The ADM mass and the electric and magnetic charges of this black hole
solutions are given by
\eqn
\arr{l}
\frac{2M}{A_{D-2}}=(D-2)\mu+(D-3){\displaystyle \sum_{i=1}^{n}}\a_{i}. 
\\\\
\left(\frac{Q_{i}}{(D-3)A_{D-2}}\right)^{2}=
\a_{i}\left(\mu+\a_{i}\right)\sin^2{\zeta_{i}},
\\\\
\left(\frac{P_{i}}{(D-3)A_{D-2}}\right)^{2}=
\a_{i}\left(\mu+\a_{i}\right)\cos^2{\zeta_{i}},
\earr
\eeqn
where $\cos{\zeta_{i}}=0$ or $\sin{\zeta_{i}}=0$ when the
corresponding electric or magnetic charge does not originate from the
eleven-dimensional $(2\subset 5)$-brane. Labelling all
the charges by $Q_{i}$ (electric, magnetic and KK in origin) and
defining for the originally ($2\subset 5$)-brane case an
electromagnetic charge $Q_{i}$ by $Q_{i}^{2}=P_{2\subset 5,i}^{2}+
P_{2\subset 5,i}^{2}$, the ADM mass can be written as
\eqn
\frac{2M}{(D-3)A_{D-2}}=\lambda\mu+{\displaystyle \sum_{i=1}^{n}}
\sqrt{\left(\frac{\mu}{2}\right)^{2}+
  \left(\frac{Q_{i}}{(D-3)A_{D-2}}\right)^{2}},
\eeqn
where $\lambda=\frac{D-2}{D-3}-\frac{n}{2}$ \cite{ct}.

We now write thermodynamical formulae. Let us start by calculating the
Hawking temperature. Write the $g_{tt}$ and $g_{rr}$ metric elements
as $g_{tt}=-M(r)f$ and $g_{rr}=L(r)f^{-1}$. We then have
$\frac{L(r)}{M(r)}={\displaystyle
  \prod_{i=1}^{n}}H_{i}(r)$. Performing the coordinate transformation
$\rho ^{2}=f(r)$, analytically continuing to Euclidean spacetime  and
examining the behaviour of the metric (4.1) in the vicinity of the
horizon $r=\mu^{\frac{1}{D-3}}$, this temperature is seen to be
\eqn
T_{H}=\frac{D-3}{4\pi}\mu^{-\frac{1}{D-3}+\frac{n}{2}}
\prod_{i=1}^{n}\left(\mu+\a_{i}\right)^{-\frac{1}{2}}.
\eeqn
We can write the Hawking temperature as a function of the physical
charges $Q_{i}$
\eqn
T_{H}=\frac{D-3}{4\pi}\mu^{-\frac{1}{D-3}+\frac{n}{2}}
\prod_{i=1}^{n}\left(\frac{\mu}{2}+\sqrt{\left(\frac{\mu}{2}\right)^{2}
    +\left(\frac{Q_{i}}{(D-3)A_{D-2}}\right)^{2}}\right)^{-\frac{1}{2}}.
\eeqn
In the extremal limit $\mu\rightarrow 0$ we have
\eqn
T_{H}\rightarrow\frac{D-3}{4\pi}\mu^{-\frac{1}{D-3}+\frac{n}{2}}
\prod_{i=1}^{n}\left(\frac{|Q_{i}|}{(D-3)A_{D-2}}\right)^{-\frac{1}{2}}.
\eeqn
For $n=1$ and $D=4$ the temperature diverges
as $\mu\rightarrow 0$. For $n=2$ and $D=4$, or $n=1$ and $D=5$ we have
$T_{H}\rightarrow \frac{D-3}{4\pi}{\displaystyle
  \prod_{i=1}^{n}}\left(\frac{|Q_{i}|}{(D-3)A_{D-2}}\right)^{-\frac{1}{2}}$. 
For $n>2$ and $D=4$,  $n>1$ and $D=5$, or $6\le D\le 10$ the temperature
converges to zero.\footnote{The $Q-M$, $T-Q$ and $T-M$ plots for the
  case $n=1$ were given in \cite{cp} in the context of KK electrically
  charged solutions. The case $n=2$ and $D=4$ exhibits the same
  features as the the case $n=1$ and $D=5$. All the other cases are
  similar to the cases with $n=1$ and $6\le D\le 10$ (see \cite{klopp} for
  some $3D$ plots).}

The Bekenstein-Hawking entropy of the black hole solutions described
by (4.1) is given by $\frac{A_{H}}{4G_{D}}$, where $A_{H}$ is the
horizon area, $G_{D}$ the $D$-dimensional Newton's constant and in our
units $4G_{D}=(2\pi )^{-1}$.\footnote{The Newton's constant in 11
and $D$ dimensions are related by $G_{11}=G_{D}L^{p}$, where $L$ is the
length of each of the $p$ compact directions. We have taken
$4G_{11}=(2\pi )^{-1}$ and $L=1$.} This entropy is given
by
\eqn
S_{BH}=2\pi A_{D-2}\mu^{\lambda}\prod_{i=1}^{n}
\left(\mu +\a_{i}\right)^{\frac{1}{2}}.
\eeqn
In terms of the
physical charges it is given by \cite{ct}
\eqn
S_{BH}=2\pi A_{D-2}\mu^{\lambda}
\prod_{i=1}^{n}\left(\frac{\mu}{2}+\sqrt{\left(\frac{\mu}{2}\right)^{2}
    +\left(\frac{Q_{i}}{(D-3)A_{D-2}}\right)^{2}}\right)^{\frac{1}{2}}.
\eeqn
In the extremal limit we have
\eqn
S_{BH}\rightarrow 2\pi A_{D-2} \mu^{\lambda}\prod_{i=1}^{n}
\left(\frac{|Q_{i}|}{(D-3)A_{D-2}}\right)^{\frac{1}{2}}.
\eeqn
Extremal black holes with finite entropy are those for which $\lambda
=0$, i.e. $D=4$ and $n=4$, $D=5$ and $n=3$ \cite{dlp}.

We now write the mass formula for our black hole solutions. Defining
$\Phi_{H}^{i}=
\frac{1}{2}\frac{Q_{i}}{(D-3)A_{D-2}}\frac{1}{\mu +\a_{i}}$ to be the
horizon electromagnetic potential due to the i-th brane (or the
originally i-th ($2\subset 5$)-brane), we have
\eqn
M=\frac{D-2}{D-3}T_{H}S_{BH}+\sum_{i=1}^{n}\Phi_{H}^{i}Q_{i}.
\eeqn
The differential form of this equation gives the first law of
thermodynamics
\eqn
dM=T_{H}dS_{BH}+\sum_{i=1}^{n}\Phi_{H}^{i}dQ_{i}.
\eeqn
In the cases where the charge $Q_{i}$ arises from the $(2\subset
5)$-brane we can have $dQ_{i}=0$ ($dM=0$ and $dS=0$) while changing
the composite branes electric and magnetic charges. It corresponds to
a rotation of the electromagnetic charges.

To conclude, we have generalised all known supersymmetric M-branes to
the corresponding black configurations. We also wrote the
thermodynamical formulae that follows from the simple rules to
construct the black composite M-branes. This way we have extended the
previous work of Duff, L\"{u} and Pope \cite{dlp} and of Cveti\v c and
Tseytlin \cite{ct}.

\section*{Acknowledgements}

The author is grateful to M.J.Perry for introducing him into the
M-theory subject and acknowledges the financial support of JNICT
(Portugal) under programme PRAXIS XXI.

\end{document}